\documentclass[sigconf,screen]{acmart}

\AtBeginDocument{%
  
}

\setcopyright{cc}
\setcctype{by}
\acmDOI{10.1145/3844135.3845866}
\acmYear{2026}
\copyrightyear{2026}
\acmISBN{979-8-4007-2999-7/2026/10}
\acmConference[ATIQSER '26]{Proceedings of the 1st International Workshop on Automated Techniques for Integrity and Quality in Software-Engineering Research}{October 12--16, 2026}{Munich, Germany}
\acmBooktitle{Proceedings of the 1st International Workshop on Automated Techniques for Integrity and Quality in Software-Engineering Research (ATIQSER '26), October 12--16, 2026, Munich, Germany}
\acmSubmissionID{asews26atiqsermain-p3-p}
\received{2026-08-16}
\received[accepted]{2026-09-01}
\newcommand{\stageOneParagraphs}{300}
\newcommand{\stageOneSentences}{1{,}703}
\newcommand{\stageOneGoldCitances}{578}
\newcommand{\drThreeFOne}{0.990}

\newcommand{\stageOnePdfRecall}{0.93}
\newcommand{\stageOneRegexPrec}{0.980}
\newcommand{\stageOneRegexFp}{12}
\newcommand{\stageOneRegexRecall}{1.000}
\newcommand{\stageOneSentFOne}{0.991}
\newcommand{\stageOneSentPrec}{0.990}
\newcommand{\stageOneSentRecall}{0.993}
\newcommand{\stageOneSpanFOne}{0.989}
\newcommand{\stageOneSpanPrec}{0.986}
\newcommand{\stageOneSpanRecall}{0.991}

\newcommand{\stageTwoSourcePapers}{23}
\newcommand{\stageTwoRefs}{992}
\newcommand{\stageTwoResolved}{571}
\newcommand{\drFiveResolution}{0.576}
\newcommand{\stageTwoNewRate}{0.90}
\newcommand{\stageTwoOldRate}{0.45}
\newcommand{\stageTwoIdRate}{0.93}
\newcommand{\stageTwoNoIdRate}{0.45}
\newcommand{\stageTwoMlRefs}{275}
\newcommand{\stageTwoPhysicsRefs}{119}
\newcommand{\stageTwoMathRefs}{140}
\newcommand{\stageTwoBiomedRefs}{458}
\newcommand{\stageTwoBiomedBefore}{0.21}
\newcommand{\stageTwoBiomedRate}{0.35}
\newcommand{\stageTwoMathRate}{0.52}
\newcommand{\stageTwoMissedOa}{113}
\newcommand{\stageTwoMlRate}{0.86}
\newcommand{\stageTwoNoFreeCopy}{231}
\newcommand{\stageTwoPhysicsRate}{0.86}
\newcommand{\stageTwoPrecision}{0.93}
\newcommand{\stageTwoUnjudged}{77}
\newcommand{\stageTwoUnresolved}{421}

\newcommand{\drFourSciFactScored}{64}
\newcommand{\drFourCitIntCitances}{80}
\newcommand{\drFourCitIntScored}{65}
\newcommand{\drFourHitAbstract}{0.94}
\newcommand{\drFourHitBaseline}{0.52}
\newcommand{\drFourHitLlmLocTerra}{0.68}
\newcommand{\drFourHitLlmLocTerraSciFact}{0.98}
\newcommand{\drFourHitReranker}{0.69}
\newcommand{\drFourHitSmallReranker}{0.60}
\newcommand{\drFourHitTopTen}{0.75}

\newcommand{\drFourMrrBaseline}{0.38}
\newcommand{\drFourMrrLlmLocTerra}{0.52}
\newcommand{\drFourMrrLlmLocTerraSciFact}{0.96}
\newcommand{\drFourMrrReranker}{0.52}
\newcommand{\drFourMrrSmallReranker}{0.42}
\newcommand{\drFourRecallBaseline}{0.36}
\newcommand{\drFourRecallReranker}{0.49}
\newcommand{\drFourRecallSmallReranker}{0.43}

\newcommand{\stageFourSweepN}{100}
\newcommand{\drOneVerdictSciFact}{0.91}
\newcommand{\drOneVerdictCitInt}{0.72}
\newcommand{\stageFourCitIntMacroF}{0.62}
\newcommand{\stageFourCitIntStrict}{0.32}
\newcommand{\stageFourModelSweepHigh}{0.65}
\newcommand{\stageFourModelSweepLow}{0.59}
\newcommand{\stageFourNemotronGold}{0.64}
\newcommand{\stageFourNemotronLocated}{0.47}
\newcommand{\stageFourSciCiteVal}{0.66}
\newcommand{\stageFourSciCiteValMacroF}{0.43}
\newcommand{\stageFourSciCiteValStrict}{0.39}
\newcommand{\stageFourSciFactMacroF}{0.89}
\newcommand{\stageFourSciFactStrict}{0.65}
\newcommand{\stageFourSCitance}{0.78}
\newcommand{\stageFourSCitanceMacroF}{0.80}
\newcommand{\stageFourSCitanceStrict}{0.60}
\newcommand{\stageFourSweepMiniGold}{0.62}
\newcommand{\stageFourSweepMiniLocated}{0.52}
\newcommand{\stageFourSweepNanoHighLocated}{0.58}
\newcommand{\stageFourSweepNanoLowLocated}{0.54}
\newcommand{\stageFourSweepTerraGold}{0.72}
\newcommand{\stageFourSweepTerraLocated}{0.62}
\newcommand{\stageFourSweepTerraLowGold}{0.72}
\newcommand{\stageFourSweepTerraLowLocated}{0.59}
\newcommand{\stageFourSweepTerraMedGold}{0.73}
\newcommand{\stageFourSweepTerraMedLocated}{0.64}

\newcommand{\annotReports}{8}
\newcommand{\annotAttrAcc}{0.86}
\newcommand{\annotGoldSupported}{118}
\newcommand{\annotKept}{145}
\newcommand{\annotRows}{153}
\newcommand{\annotStageFourFourAcc}{0.71}

\newcommand{\annotStageOneFOne}{0.97}
\newcommand{\annotStageOnePrec}{0.95}
\newcommand{\annotStageOneRecall}{1.00}
\newcommand{\annotStageThreeFOne}{0.67}
\newcommand{\annotStageThreePrec}{0.53}
\newcommand{\annotStageThreeRecall}{0.91}
\newcommand{\annotStageTwoAccKept}{1.00}
\newcommand{\annotSupToPartial}{25}

\newcommand{\drOneRefsMin}{23}
\newcommand{\drOneRefsMax}{101}
\newcommand{\drOneRefsMedian}{39}
\newcommand{\drOneClaimsMin}{28}
\newcommand{\drOneClaimsMax}{174}
\newcommand{\drOneClaimsMedian}{60.5}
\newcommand{\drOneRuntimeMedianSec}{634}
\newcommand{\drOneRuntimeMedianMin}{11}

\newcommand{\drOneStageTwoShareLow}{0.35}
\newcommand{\drOneStageTwoShareHigh}{0.91}
\newcommand{\drOneExtractVerdictMaxMin}{6.3}

\newcommand{\surveyCOneMed}{6}
\newcommand{\surveyEFiveMed}{7}
\newcommand{\surveyEFourMed}{5}
\newcommand{\surveyEOneMed}{6}

\newcommand{\surveyETwoMed}{6}
\newcommand{\surveyN}{11}
\newcommand{\surveyPuMedian}{6}

\usepackage{microtype}
\usepackage{balance}
\usepackage{booktabs}
\usepackage{tabularx}
\usepackage{subcaption}
\usepackage{tikz}
\usepackage{xspace}
\usepackage{xfp}
\usepackage{cleveref}
\usepackage{svg}
\usetikzlibrary{shapes,arrows.meta,positioning,decorations.pathreplacing}

\definecolor{TUMLightGray}{RGB}{240,240,240}
\definecolor{TUMAccentLightBlue}{RGB}{100,160,220}
\definecolor{TUMAccentOrange}{RGB}{230,140,50}
\definecolor{TUMAccentGreen}{RGB}{80,180,100}

\definecolor{EvalSlate}{RGB}{110,120,150}     % stage boxes (was LightBlue)
\definecolor{EvalPurple}{RGB}{150,100,190}    % benchmark boxes (was Orange)
\definecolor{EvalTeal}{RGB}{60,170,175}       % sample boxes (was Green)

\newcommand{\toPercent}[1]{\fpeval{round(#1 * 100, 1)}\%}
\newcommand{\toolname}{RefVerifier\xspace}

\begin{document}

\title{\toolname: Semi-Automated Reference Claim Verification for Scientific Manuscripts}

\author{Stefania Mocan}
\orcid{0009-0005-7587-3193}
\affiliation{%
  \institution{Technical University of Munich}
  \city{Munich}
  \country{Germany}
}

\author{Florian Angermeir}
\orcid{0000-0001-7903-8236}
\affiliation{%
  \institution{fortiss}
  \city{Munich}
  \country{Germany}
}
\affiliation{%
  \institution{Blekinge Institute of Technology}
  \city{Karlskrona}
  \country{Sweden}
}

\author{Mark Kreitz}
\orcid{0009-0001-8282-8930}
\affiliation{
  \institution{University of the Bundeswehr Munich}
  \city{Munich}
  \country{Germany}
}
\affiliation{
  \institution{Blekinge Institute of Technoloy}
  \city{Karlskrona}
  \country{Sweden}
}

\renewcommand{\shortauthors}{Mocan, Angermeir, Kreitz}

\begin{abstract}
As software engineering research submission counts surge, peer reviewers face severe time constraints, making systematic verification of citation-supported claims prohibitively expensive. Consequently, unsubstantiated claims and semantic drift can propagate undetected across scientific literature. Existing approaches such as fact-checking and retrieval-augmented generation tools operate on open-domain web data or evaluate claims in isolation without processing complete manuscripts. 
To address this gap, we present \textbf{\toolname}, a semi-automated, citation-bounded reference verification prototype designed to support in academic peer review. \toolname extracts citation-bearing sentences from manuscripts, checks bibliography metadata against scholarly databases, resolves references to full-text open-access PDFs, localizes relevant evidence passages, and generates verdicts with natural language explanations.

Evaluating \toolname{} on public benchmarks shows
claim detection at an F1 score of \drThreeFOne{}, open-access resolution of
\toPercent{\drFiveResolution{}} of references, and evidence localization with a
hit rate of \toPercent{\drFourHitLlmLocTerraSciFact{}} on abstracts and
\toPercent{\drFourHitLlmLocTerra{}} on complete cited papers. In an end-to-end test with eight manuscripts, \toolname achieves a verdict accuracy of \toPercent{\annotStageFourFourAcc{}}. By automating document retrieval and evidence localization while preserving reviewer oversight, \toolname provides first indicators for the feasibility of semi-automated integrity checks in scholarly publishing.

\end{abstract}




\maketitle

\section{Introduction}

Scientific progress relies on citations that establish an auditable evidence trail for asserted claims~\cite{Ngatuvai2021}. Citations attribute prior findings~\cite{Panjaitan2024}, prevent plagiarism~\cite{Helgesson2015}, and serve as a cornerstone of peer review quality control~\cite{Neville2012, Divecha2023}. 

In practice, however, reviewers rarely verify whether cited sources actually support the accompanying assertions due to strict submission deadlines and heavy workloads~\cite{Sovacool2022}. As a result, inaccurate citations, overstatements, and unsubstantiated claims pass review and propagate through the literature. Greenberg demonstrated how an unsubstantiated claim regarding muscle damage propagated through a citation network of 242 papers, gaining unwarranted academic authority simply through repeated citation without underlying data~\cite{Greenberg2009}. Generative Artificial Intelligence (AI) tools exacerbate this risk by generating fluent text with hallucinated references or superficial citation placeholders~\cite{Camp2025, Walters2023}.

Existing verification tools offer limited support for peer review. Open-domain fact-checking systems (e.g., FEVER~\cite{Thorne2018}) assess claim veracity against broad corpora rather than validating claims strictly against author-designated references. Traditional reference managers perform only surface-level metadata checking (e.g., DOI formatting, volume consistency)~\cite{Heibi2025}. Meanwhile, specialized citation tools such as SemanticCite~\cite{Haan2025} evaluate isolated, manually supplied claims rather than parsing complete manuscripts.

To bridge this research gap, we present \textbf{\toolname}, a semi-automated reference claim verification system for scientific manu\-scripts. \toolname restricts its evidence strictly to the references cited in the manuscript, executing a four-stage pipeline: (1) deterministic extraction and LLM enrichment of citation-bound claims, (2) automated reference metadata checking and multi-source open-access paper retrieval, (3) passage-level evidence localization within full cited texts, and (4) verdict prediction with explanations. The reviewer maintains control over the final assessment, guided by an interactive split-pane interface.

\textbf{Contributions:}
\begin{itemize}
    \item We formulate the task of manuscript-level, citation-bounded reference verification and identify the research gap in current peer-review automation tools.
    \item We introduce the architecture and design of \textbf{\toolname}, a proof-of-concept tool to support reviewer in reference claim verification.
    \item We build a small end-to-end testing dataset comprising eight manuscripts.
    \item We conduct an empirical quantitative evaluation across established public benchmarks (CiteWorth, SciFact, SCitance, Citation-Integrity, SciCiteVal) and a qualitative technology acceptance evaluation with 11 experienced reviewers.
\end{itemize}

\section{Related Work}
\label{sec:background}
We present the foundational definitions, relevant datasets and the research gap.

\subsection{Foundational Definitions}
\label{subsec:definitions}

We distinguish three core concepts in reference verification:
\begin{itemize}
    \item \textbf{Claim}: A verifiable proposition expressing a finding about a scientific entity or process~\cite{Wadden2020}.
    \item \textbf{Citance}: A sentence containing an inline citation that asserts a statement regarding the cited work~\cite{Nakov2004}. The citance serves as our primary unit of verification.
    \item \textbf{Fact}: A claim whose evidential support is sufficiently strong to justify treating it as true~\cite{Nissen2016}.
\end{itemize}

While open-domain \emph{fact-checking} evaluates claims against external knowledge bases (e.g., Wikipedia or web crawls) to issue global truth verdicts~\cite{Vladika2024}, \emph{reference claim verification} evaluates whether an author-designated source substantiates the asserted claim, regardless of external consensus~\cite{Vladika2023}. Similarly, Retrieval-Augmented Generation (RAG) and Question Answering (QA) over scientific papers (e.g., QASPER~\cite{Dasigi2021}, QASA~\cite{Lee2023}) optimize for informativeness rather than verifying claim-evidence alignment.

\subsection{Benchmark Datasets}
General-domain fact-checking benchmarks such as FEVER~\cite{Thorne2018} and  AVeriTeC~\cite{Schlichtkrull2023} use open retrieval over large web or Wikipedia corpora. Abstract-level datasets (SciFact~\cite{Wadden2020}, SCitance~\cite{Alvarez2024}) evaluate whether scientific claims are supported, but limit verification to paper abstracts. At the full-paper level, Citation-Integrity~\cite{Sarol2024} provides ground-truth evidence spans for biomedical citations, while SciCiteVal~\cite{Liu2026} constructs incorrect citation contexts through controlled distortions rather than using naturally occurring miscitations. SemanticCite~\cite{Haan2025} performs LLM-based citation checking against full reference documents but operates on supplied citation text rather than parsing complete manuscripts. In summary, existing benchmarks lack citation-bounded verification, rely on isolated abstracts, or assume claims are pre-extracted, establishing the gap that \toolname addresses. The full list and comparison of established datasets is provided in the online material~\cite{online_material}.

\subsection{Research Gap}
\label{sec:gap}
Existing approaches suffer from four fundamental limitations:
1. \textbf{Lack of citation-bounded verification}: Open-domain systems search external corpora rather than restricting verification to author-cited sources.
2. \textbf{Unextracted claims}: Most tools assume pre-packaged claims and cannot identify citation-bearing claims directly from complete manuscripts.
3. \textbf{Fragmented workflows}: Prior work addresses isolated sub-tasks (e.g., citation-worthiness detection via CiteWorth~\cite{Wright2021}) without offering an end-to-end workflow from manuscript parsing to verdict presentation.
4. \textbf{Limited full-paper support}: Datasets rely heavily on abstracts or isolated passages rather than complete multi-page documents.

\toolname addresses this research gap by delivering an end-to-end, citation-bounded reference verification tool operating on full manuscripts.

\section{Prototype Design}
\label{sec:implementation}

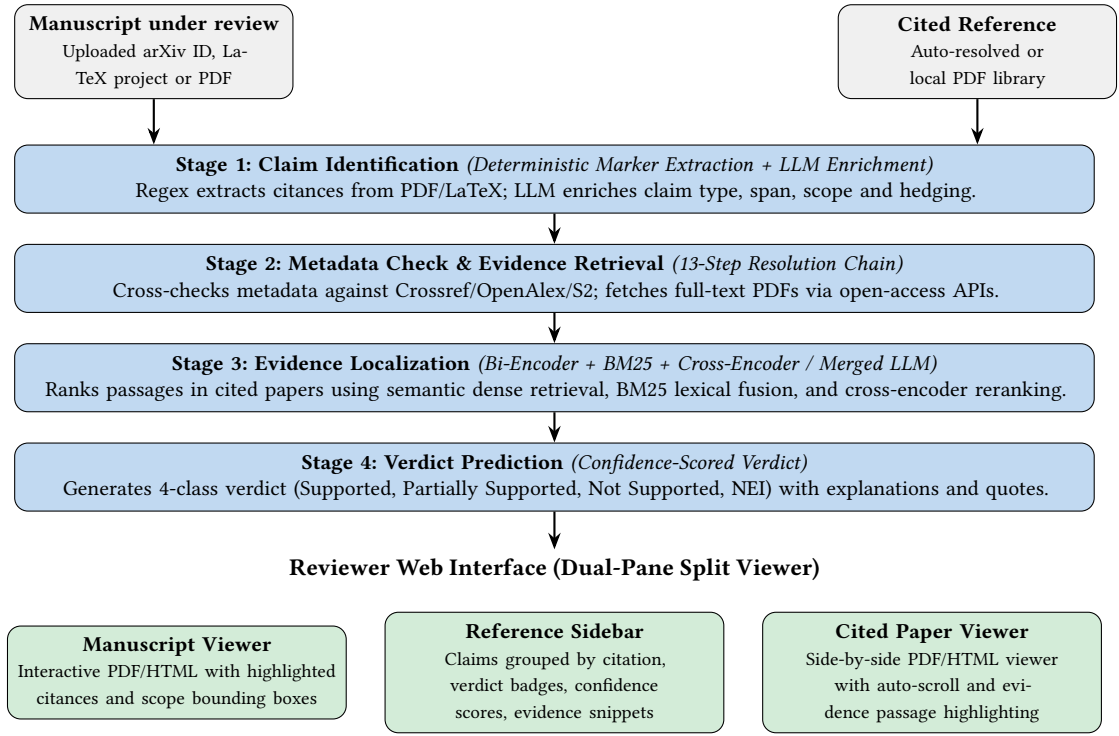
\begin{figure*}[t]
\centering
\begin{tikzpicture}[
    font=\small,
    node distance=5mm,
    arrow/.style={-{Stealth[length=2.2mm]}, thick},
    sidearrow/.style={-{Stealth[length=2mm]}, thick, dashed},
    box/.style={draw, rounded corners=5pt, align=center, inner sep=4pt},
    input/.style={box, fill=TUMLightGray, text width=3.4cm, minimum height=8mm},
    stage/.style={box, fill=TUMAccentLightBlue!40, text width=14cm, minimum height=9mm},
    support/.style={box, dashed, fill=TUMAccentOrange!25, text width=3.2cm, minimum height=9mm},
    pane/.style={box, fill=TUMAccentGreen!25, text width=4.2cm, minimum height=11mm}
]

% Inputs
\node[input] (man) {\textbf{Manuscript under review}\\\footnotesize Uploaded arXiv ID, LaTeX project or PDF};
\node[input, right=7.2cm of man] (refs) {\textbf{Cited Reference}\\\footnotesize Auto-resolved or local PDF library};

% Stage 1
\node[stage, below=6mm of man.south west, anchor=north west] (s1) {%
  \textbf{Stage 1: Claim Identification} \textit{(Deterministic Marker Extraction + LLM Enrichment)}\\%
  Regex extracts citances from PDF/LaTeX; LLM enriches claim type, span, scope and hedging.%
};
\draw[arrow] (man.south) -- (man.south |- s1.north);
\draw[arrow] (refs.south) -- (refs.south |- s1.north);

% Stage 2
\node[stage, below=4mm of s1] (s2) {%
  \textbf{Stage 2: Metadata Check \& Evidence Retrieval} \textit{(13-Step Resolution Chain)}\\%
  Cross-checks metadata against Crossref/OpenAlex/S2; fetches full-text PDFs via open-access APIs.%
};
\draw[arrow] (s1) -- (s2);

% Stage 3
\node[stage, below=4mm of s2] (s3) {%
  \textbf{Stage 3: Evidence Localization} \textit{(Bi-Encoder + BM25 + Cross-Encoder / Merged LLM)}\\%
  Ranks passages in cited papers using semantic dense retrieval, BM25 lexical fusion, and cross-encoder reranking.%
};
\draw[arrow] (s2) -- (s3);

% Stage 4
\node[stage, below=4mm of s3] (s4) {%
  \textbf{Stage 4: Verdict Prediction} \textit{(Confidence-Scored Verdict)}\\%
  Generates 4-class verdict (Supported, Partially Supported, Not Supported, NEI) with explanations and quotes.%
};
\draw[arrow] (s3) -- (s4);

% UI
\node[font=\bfseries, below=5mm of s4] (iface) {Reviewer Web Interface (Dual-Pane Split Viewer)};
\draw[arrow] (s4.south) -- (iface.north);

\node[pane, below=3mm of iface] (p2) {\textbf{Reference Sidebar}\\\footnotesize Claims grouped by citation, verdict badges, confidence scores, evidence snippets};
\node[pane, left=5mm of p2] (p1) {\textbf{Manuscript Viewer}\\\footnotesize Interactive PDF/HTML with highlighted citances and scope bounding boxes};
\node[pane, right=5mm of p2] (p3) {\textbf{Cited Paper Viewer}\\\footnotesize Side-by-side PDF/HTML viewer with auto-scroll and evidence passage highlighting};

\end{tikzpicture}
\caption{Overview of the \toolname four-stage verification pipeline and interactive reviewer web interface.}
\label{fig:pipeline_architecture}
\end{figure*}

\toolname is implemented as a web application featuring a Python FastAPI backend and a React 19 single-page frontend. Figure~\ref{fig:pipeline_architecture} illustrates the core architecture across its four sequential stages.

\subsection*{Stage 1: Claim Identification}
\label{sec:impl-extractor}

Claim identification extracts citation-bearing sentences and enriches them with structural metadata through a hybrid deterministic-LLM approach:

\textbf{Deterministic Detection}: Inline citation markers (e.g., \texttt{[12]}, \texttt{(Smith et al., 2020)}) are identified in extracted PDF text using PyMuPDF (\texttt{fitz}) or directly parsed from LaTeX citation keys. If the PDF carries no extractable text layer, Tesseract OCR is invoked as a fallback.

\textbf{LLM Enrichment}: Detected citances are sent to the LLM (currently \texttt{gpt-5.6-terra}) in fixed-size batches. The model enriches each citance with:
   i) \textbf{Claim Type}: Categorized into seven rubrics based on Magnusson and Friedman~\cite{Magnusson2021} (Causal, Comparative, Predictive, Proportional, Statistical) and extended by two further types (Background, Naming/Attribution).
   ii) \textbf{Hedge Level}: Deterministic regex scoring flags uncertainty markers (e.g., \emph{"suggests"}, \emph{"may indicate"}), relaxing verdict criteria so cautious assertions are not penalized.
   iii) \textbf{Span Split}: In a sentence with multiple citations, each citation is assigned only its corresponding verbatim span (e.g., \emph{X achieves 95\% [1], while Y runs in 10ms [2]''}, where the verbatim span for [1] is \emph{``X achieves 95\%''} and for [2] is \emph{``Y runs in 10ms''}). Citations supporting the same proposition receive the same claim and verbatim span.
   iv) \textbf{Citation Scope}: The model records any lead-in sentences before the citation marker and continuation sentences after it that are supported by the same source. The citation span and its surrounding scope are verified together.
   v) \textbf{Ambiguity Handling}: Dangling pronouns or missing referents raise an ambiguity flag, triggering an LLM rewrite with explicit interpretation choices.

\subsection*{Stage 2: Metadata Check \& Evidence Retrieval}
\label{sec:impl-retriever}

Before semantic verification, a meta-data checker validates reference existence. Bibliography entries are parsed into structured fields (authors, title, year, venue, DOI) and cross-checked against Semantic Scholar, Crossref, and OpenAlex. References are flagged as \emph{validated}, \emph{unvalidatable} (e.g., unpublished manuscripts), or \emph{suspicious} (e.g., title/author contradictions indicative of hallucinated citations).

Next, the retriever resolves each citation key to a full-text PDF via a multistep resolution chain: The resolver first checks a local cache and user uploads, falls back to direct ID lookups (arXiv, ACL), and sequentially queries open-access APIs like Semantic Scholar, Crossref, OpenAlex, Unpaywall, and Europe PMC before attempting a free-text search on Google Scholar. The resolver stops at the first valid PDF. Confirmed PDFs undergo a deterministic "right-paper" check matching embedded metadata and first-page titles against the reference.

\subsection*{Stage 3: Evidence Localization}
\label{sec:impl-embedder}

Evidence localization identifies passages within the cited paper that substantiate the claim. \toolname implements two localizer modes:

\textbf{Two-Stage Retriever}: The cited paper is divided into overlapping 3-sentence windows (5 sentences for causal/predictive claims). Ranking proceeds via:

\emph{Semantic Ranking}: Bi-encoder \texttt{all-MiniLM-L6-v2} computes cosine similarity between claim and passage embeddings.

\emph{Lexical Fusion}: Okapi BM25 scores keyword overlap. Reciprocal Rank Fusion (RRF) merges dense and sparse rankings.

\emph{Cross-Encoder Reranking}: A cross-encoder (\texttt{bge-reranker-v2-m3}) reranks top candidates, scoring relevance confidence.

\textbf{Merged LLM Mode (Deployed Default)}: A single LLM call processes the full text of the cited paper, directly identifying supporting passages and generating the verdicts (see Stage 4) at once, bypassing separate retrieval models and effectively combining stage three and four into a single one. 

\subsection*{Stage 4: Verdict Prediction}
\label{sec:impl-verdict}

The verdict engine evaluates localized evidence against the claim, returning a four-class verdict:
\begin{itemize}
    \item \textbf{Supported}: Evidence directly confirms the claim assertion.
    \item \textbf{Partially Supported}: Evidence partially substantiates the claim but reveals minor overstatements, missing nuances, or scope shifts.
    \item \textbf{Not Supported}: Evidence directly refutes the claim or provides contradictory data.
    \item \textbf{Not Enough Information (NEI)}: The tool could not confidently make a verdict based on the cited document.
\end{itemize}

Prompts incorporate claim-type rubrics and hedge adjustments, returning a natural language explanation.

\subsection*{Web Interface}
The main page contains an upload pane where the user can choose between the three input sources: a manuscript PDF, an arXiv ID or URL, or a LaTeX project. Once the analysis starts, a progress bar tracks the processing stages. When all stages are complete, the user is forwarded to the frontend split view for reviewing the results. The left pane presents the manuscript under analysis. Clicking a highlighted citance in the manuscript pane auto-scrolls the reference sidebar to its claim card, displaying verdict badges and explanations. Clicking an evidence snippet opens the cited paper pane and auto-scrolls the cited-paper pane directly to the highlighted passage in the resolved PDF.

\section{Evaluation}
\label{sec:evaluation}

We evaluated \toolname quantitatively through benchmark analyses and qualitatively through a technology acceptance survey.

\subsection{Quantitative Evaluation}

\begin{figure}[htpb]
  \centering
  \begin{tikzpicture}[
      font=\scriptsize,
      node distance=2mm,
      arrow/.style={-{Stealth[length=1.5mm]}, thick},
      stage/.style={
          draw, rounded corners=3pt, fill=EvalSlate!35,
          align=center, text width=2.55cm, minimum height=7mm
      },
      benchmark/.style={
          draw, rounded corners=3pt, fill=EvalPurple!25,
          align=center, text width=1.6cm, minimum height=6mm
      },
      sample/.style={
          draw, rounded corners=3pt, fill=EvalTeal!30,
          align=center, text width=1.6cm, minimum height=6mm
      }
  ]

  \node[stage] (s1) {S1 Claim identification};
  \node[stage, below=of s1] (s2) {S2 Reference retrieval};
  \node[stage, below=of s2] (s3) {S3 Evidence localization};
  \node[stage, below=of s3] (s4) {S4 Verdict prediction};

  \draw[arrow] (s1) -- (s2);
  \draw[arrow] (s2) -- (s3);
  \draw[arrow] (s3) -- (s4);

  \node[benchmark, left=2mm of s1] (citeworth)
      {CiteWorth and\\arXiv PDFs};
  \node[sample, left=2mm of s2] (retrieval)
      {\stageTwoSourcePapers{} papers\\in four domains\\(\stageTwoRefs{} references)};
  \node[benchmark, left=2mm of s3] (localization)
      {SciFact and\\Citation-Integrity};
  \node[benchmark, left=2mm of s4] (verdict)
      {SciFact, SCitance,\\SciCiteVal,\\Citation-Integrity};
  \node[sample, right=2mm of s4] (e2e)
      {Eight\\manuscripts\\across S1--S4};

  \draw[arrow] (citeworth) -- (s1);
  \draw[arrow] (retrieval) -- (s2);
  \draw[arrow] (localization) -- (s3);
  \draw[arrow] (e2e) -- (s4);
  \draw[arrow] (verdict) -- (s4);

  \end{tikzpicture}
  \caption{Quantitative evaluation coverage. Purple boxes denote public benchmarks. Teal boxes denote the authors' sampled-reference and end-to-end studies.}
  \label{fig:evaluation_coverage}
  \end{figure}
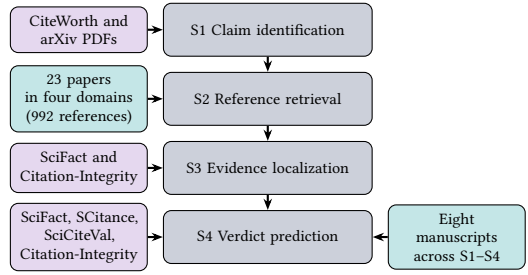

For the quantitative evaluation we employed five public datasets (CiteWorth~\cite{Wright2021}, SciFact~\cite{Wadden2020}, SCitance~\cite{Alvarez2024}, SciCiteVal~\cite{Liu2026}, Citation-Integrity~\cite{Sarol2024}) across the four pipeline stages. These datasets differ in task formulation, evidence scope, and label schemes, and therefore provide complementary evaluation settings. We additionally tested \toolname{} on a purpose-build end-to-end dataset of manuscripts. \Cref{fig:evaluation_coverage} provides a high-level overview of the quantitative evaluation strategy.

\paragraph*{Stage 1: Claim Identification}
\toolname extracts claims at F1 \drThreeFOne{} on CiteWorth, at high precision and recall.

We evaluated claim detection on \stageOneParagraphs{} test paragraphs (\stageOneSentences{} sentences, \stageOneGoldCitances{} gold citances) from CiteWorth~\cite{Wright2021}. As shown in \Cref{tab:stage1_results}, deterministic regex detection achieves an F1 score of \drThreeFOne{}, with precision \stageOneRegexPrec{} and recall \stageOneRegexRecall{}. This corresponds to \stageOneRegexFp{} false positives across and no sentence-level false negatives. To evaluate the performance on PDFs, we choose ten arXiv published manuscripts and evaluated the extraction on the arXiv-rendered PDFs against the corresponding \texttt{\textbackslash cite} commands in the LaTeX source. Detection recall reached \stageOnePdfRecall{}.

\begin{table}[hbtp]
\caption{Stage 1 claim detection performance on CiteWorth data.}
\label{tab:stage1_results}
\small
\setlength{\tabcolsep}{3pt}
\begin{tabularx}{\linewidth}{X c c c}
\toprule
\textbf{Evaluation Level} & \textbf{Precision} & \textbf{Recall} & \textbf{F1 Score} \\
\midrule
Regex Detection Only & \stageOneRegexPrec{} & \stageOneRegexRecall{} & \drThreeFOne{} \\
Sentence-Level (End-to-End) & \stageOneSentPrec{} & \stageOneSentRecall{} & \stageOneSentFOne{} \\
Span-Level (End-to-End) & \stageOneSpanPrec{} & \stageOneSpanRecall{} & \stageOneSpanFOne{} \\
\bottomrule
\end{tabularx}
\end{table}

\paragraph*{Stage 2: Reference Retrieval Availability}
\toolname retrieves the full-text PDFs for  \toPercent{\drFiveResolution{}} of references.

We evaluated open-access PDF retrieval on \stageTwoRefs{} references cited across \stageTwoSourcePapers{} randomly selected source papers from four research domains. \toolname resolves \toPercent{\drFiveResolution{}} (\stageTwoResolved{}/\stageTwoRefs{}) of these references to full-text PDFs. As shown in \Cref{tab:stage2_availability}, resolution rates vary substantially by domain, with Medicine and Biology showing the lowest resolution rates. Adding Europe PMC and CORE integration raised biomedical availability from \toPercent{\stageTwoBiomedBefore{}} to \toPercent{\stageTwoBiomedRate{}}. A title-matching precision audit confirmed that \toPercent{\stageTwoPrecision{}} of downloaded PDFs match the cited work.
An open-access cross-check of the \stageTwoUnresolved{} unresolved references revealed that \stageTwoNoFreeCopy{} (\fpeval{round(\stageTwoNoFreeCopy{} / \stageTwoUnresolved{} * 100, 1)}\%) have no free copy anywhere on the web (paywalled), \stageTwoUnjudged{} (\fpeval{round(\stageTwoUnjudged{} / \stageTwoUnresolved{} * 100, 1)}\%) lack identifiers, and \stageTwoMissedOa{} (\fpeval{round(\stageTwoMissedOa{} / \stageTwoUnresolved{} * 100, 1)}\%) represent open-access copies missed by the chain. Reachability is strongly predicted by publication recency (\toPercent{\stageTwoNewRate{}} for 2020-2022 vs. \toPercent{\stageTwoOldRate{}} for pre-2015) and DOI/arXiv identifier presence (\toPercent{\stageTwoIdRate{}} with ID vs. \toPercent{\stageTwoNoIdRate{}} without).

\begin{table}[hbtp]
\caption{Reference resolution availability across domains (\stageTwoRefs{} references).}
\label{tab:stage2_availability}
\small
\setlength{\tabcolsep}{3pt}
\begin{tabularx}{\linewidth}{X c c}
\toprule
\textbf{Domain Cluster} & \textbf{References} & \textbf{Reference Resolution} \\
\midrule
Computer Science / ML & \stageTwoMlRefs{} & \toPercent{\stageTwoMlRate{}} \\
Physics & \stageTwoPhysicsRefs{} & \toPercent{\stageTwoPhysicsRate{}} \\
Mathematics & \stageTwoMathRefs{} & \toPercent{\stageTwoMathRate{}} \\
Biomedicine & \stageTwoBiomedRefs{} & \toPercent{\stageTwoBiomedRate{}} \\
\midrule
\textbf{Overall} & \textbf{\stageTwoRefs{}} & \textbf{\toPercent{\drFiveResolution{}}} \\
\bottomrule
\end{tabularx}
\end{table}

\paragraph*{Stage 3: Evidence Localization}

We evaluate evidence localization by measuring whether the annotated evidence appears within the top three returned passages ($k=3$ throughout). We use subsets of SciFact (abstracts, \drFourSciFactScored{} scored claims) and Citation-Integrity (full papers, \drFourCitIntCitances{} test citances, \drFourCitIntScored{} with annotated spans), and compare several retrieval and reranking configurations together with the merged LLM mode.
Localization performs better on SciFact abstracts, where the bi-encoder reaches a hit rate of \toPercent{\drFourHitAbstract{}}. Full-paper localization is more challenging. On Citation-Integrity, the bi-encoder retrieves the correct passage into its top-10 candidate pool for \toPercent{\drFourHitTopTen{}} of citances, but ranks it among the top three in only \toPercent{\drFourHitBaseline{}} of cases, indicating that ranking is a limitation. Cross-encoder reranking raises the top-3 hit rate to \toPercent{\drFourHitReranker{}} (Table~\ref{tab:localization_rerank}).

The deployed \textbf{Merged LLM Mode} (\texttt{gpt-5.6-terra}) reaches a hit rate of \textbf{\toPercent{\drFourHitLlmLocTerra{}}} on full papers and \textbf{\toPercent{\drFourHitLlmLocTerraSciFact{}}} on abstracts (MRR \drFourMrrLlmLocTerraSciFact{}) outperforming our previous approach using a bi-encoder and re-ranker. Merged mode selects evidence sentences directly from full paper text, eliminating multi-stage pipeline latency.
\begin{table}[hbtp]
\caption{Evidence localization performance on Citation-Integrity full papers ($k=3$ passages) with gpt-5.6-terra for merge LLM mode.}
\label{tab:localization_rerank}
\small
\setlength{\tabcolsep}{2pt}
\begin{tabularx}{\linewidth}{X c c c}
\toprule
\textbf{Retrieval Configuration} & \textbf{Hit Rate} & \textbf{MRR} & \textbf{Recall@3} \\
\midrule
Bi-Encoder Cosine Baseline & \drFourHitBaseline{} & \drFourMrrBaseline{} & \drFourRecallBaseline{} \\
+ Small Reranker (ms-marco) & \drFourHitSmallReranker{} & \drFourMrrSmallReranker{} & \drFourRecallSmallReranker{} \\
+ Strong Reranker (bge-m3) & \textbf{\drFourHitReranker{}} & \textbf{\drFourMrrReranker{}} & \textbf{\drFourRecallReranker{}} \\
\midrule
\textbf{Merged LLM Mode} & \textbf{\drFourHitLlmLocTerra{}} & \textbf{\drFourMrrLlmLocTerra{}} & -- \\
\bottomrule
\end{tabularx}
\end{table}

\paragraph*{Stage 4: Verdict Prediction}
We evaluate verdict prediction on ground-truth evidence, isolating the judgment step from retrieval errors. Because our rubric contains four classes while the benchmarks use three, \emph{Partially Supported} is mapped to \emph{Supported} under the lenient mapping and to \emph{Not Enough Information} under the strict mapping. 
We report accuracy and Macro F1 for the different datasets in (\Cref{tab:verdict_accuracy}).
Across the four datasets, F1 ranges from \stageFourSciCiteValMacroF{} to \stageFourSciFactMacroF{}.
This range can be explained by the differences in dataset construction and label distributions. SciFact contains clean, human-annotated contradictions. SCitance’s score is inflated by explicit negation cues in its LLM-generated refutations. Citation-Integrity combines genuine contradictions with subtler errors such as weak substantiation and oversimplification, which do not map cleanly onto our tool’s verdict classes. SciCiteVal additionally contains constructed errors, provides only short excerpts, and has no \emph{Not Enough Information} examples, which particularly lowers the Macro-F1 of our tool.

\begin{table}[hbtp]
\caption{Verdict performance on four public datasets with gpt-5.6-terra. Mappings: Lenient (L) / Strict (S).}
\label{tab:verdict_accuracy}
\small
\setlength{\tabcolsep}{2pt}
\begin{tabularx}{\linewidth}{X c c c}
\toprule
\textbf{Dataset} & \textbf{Acc. (L)} & \textbf{Acc. (S)} & \textbf{F1 (L)} \\
\midrule
SciFact & \drOneVerdictSciFact{} & \stageFourSciFactStrict{} & \stageFourSciFactMacroF{} \\
Citation-Integrity & \drOneVerdictCitInt{} & \stageFourCitIntStrict{} & \stageFourCitIntMacroF{} \\
SCitance & \stageFourSCitance{} & \stageFourSCitanceStrict{} & \stageFourSCitanceMacroF{} \\
SciCiteVal & \stageFourSciCiteVal{} & \stageFourSciCiteValStrict{} & \stageFourSciCiteValMacroF{} \\
\bottomrule
\end{tabularx}
\end{table}

During the development of \toolname, we employed small LLMs for cost-reasons. Those models are GPT-5.4-nano, GPT-5.4-mini. To provide a long-term comparable baseline we also tested the approach using an open-source model (Nemotron 3 Ultra 550B), as requested in literature~\cite{llmguidelines, angermeir2025reflections}. For the deployed and finally evaluated version we used gpt-5.6-terra.
To understand how verdict prediction accuracy scales with different underlying architectures, we evaluated the same \stageFourSweepN{} Citation-Integrity test claims across the different models and changing reasoning efforts (Table~\ref{tab:model_sweep}) for two setups: First, when handed the relevant passages from the dataset, and second, when provided with the top three passages from our earlier stages only.

For the small models, performance is rather similar to the open model, being on par with commercial ones. Nevertheless, gpt-5.6-terra performed at any reasoning level better than the small models, but also at an increase in inference cost.

\begin{table}[htbp]
\centering
\caption{Verdict accuracy by model configuration on Citation-Integrity (100 test citances, lenient accuracy). We distinguish between (1) dataset passages provided to stage 4 and (2) the top 3 passages form ealier stages provided to stage 4.}
\label{tab:model_sweep}
\small
\begin{tabular}{llcc}
\toprule
\textbf{Model} & \textbf{Effort} & \textbf{(1) Dataset} & \textbf{(2) Top 3} \\
\midrule
\texttt{gpt-5.4-nano}    & low    & \stageFourModelSweepLow{}   & \stageFourSweepNanoLowLocated{} \\
\texttt{gpt-5.4-nano}    & high   & \stageFourModelSweepHigh{}  & \stageFourSweepNanoHighLocated{} \\
\texttt{gpt-5.4-mini}    & medium & \stageFourSweepMiniGold{}   & \stageFourSweepMiniLocated{} \\
Nemotron~3 Ultra (550B)  & n/a    & \stageFourNemotronGold{}    & \stageFourNemotronLocated{} \\
\midrule
\texttt{gpt-5.6-terra}   & low    & \stageFourSweepTerraLowGold{} & \stageFourSweepTerraLowLocated{} \\
\texttt{gpt-5.6-terra}   & medium & \stageFourSweepTerraMedGold{}       & \stageFourSweepTerraMedLocated{} \\
\texttt{gpt-5.6-terra}   & high   & \textbf{\stageFourSweepTerraGold{}} & \textbf{\stageFourSweepTerraLocated{}} \\
\bottomrule
\end{tabular}
\end{table}

\paragraph*{End-to-End Manuscript Verification}

We evaluate \toolname on eight real-world scientific manuscripts comprising \annotRows{} manually annotated citance-reference rows (\annotKept{} confirmed pairs). \Cref{tab:endtoend} lists the performance on the stages. Verdicts were blindly annotated, while citance text, reference mapping, and evidence passages were refined from the proposals of \toolname, with missing elements added during review when necessary. For evidence localization, the tool typically retrieves the relevant passages while also returning additional candidates that may still be relevant but were not selected by the annotator as the strongest evidence, resulting in lower precision than recall. Most verdict disagreements reflect a cautious tendency toward \emph{Partially Supported} where the annotator selected \emph{Supported} (\annotSupToPartial{} of \annotGoldSupported{} claims).

\begin{table}[htbp]
\caption{End-to-end evaluation metrics across \annotReports{} manuscripts (\annotKept{} confirmed citance-reference pairs) with gpt-5.6-terra. P=Precision, R=Recall.}
\label{tab:endtoend}
\small
\setlength{\tabcolsep}{2pt}
\begin{tabularx}{\linewidth}{X l c}
\toprule
\textbf{Pipeline Stage} & \textbf{Evaluation Metric} & \textbf{Value} \\
\midrule
S1: Claim Detection & P / R / F1 & \annotStageOnePrec{} / \annotStageOneRecall{} / \annotStageOneFOne{} \\
S1: Claim Attribution & Attribution Accuracy & \annotAttrAcc{} \\
S2: Reference Mapping & Mapping Accuracy & \annotStageTwoAccKept{} \\
S3: Evidence Localize & P/ Micro R / F1 & \annotStageThreePrec{} / \annotStageThreeRecall{} / \annotStageThreeFOne{} \\
S4: Verdict Prediction & 4-Class Accuracy & \annotStageFourFourAcc{} \\
\bottomrule
\end{tabularx}
\end{table}

\textbf{Execution Runtime Feasibility}: Across all live manuscript runs with \drOneRefsMin{}-\drOneRefsMax{} references (\drOneRefsMedian{} median) and \drOneClaimsMin{}-\drOneClaimsMax{} claims per paper (\drOneClaimsMedian{} median)), the median end-to-end processing time was \textbf{\drOneRuntimeMedianSec{} seconds} (\drOneRuntimeMedianMin{} minutes).
Stage 2 reference retrieval dominated execution time, accounting for \toPercent{\drOneStageTwoShareLow{}}-\toPercent{\drOneStageTwoShareHigh{}} (median >65\%) of total runtime due to external API restrictions and PDF downloads. 
LLM claim extraction and verdict generation completed in at most \drOneExtractVerdictMaxMin{} minutes per paper.

\subsection{Qualitative Evaluation}
To assess qualitative acceptance, we conducted a technology acceptance survey ($N=\surveyN{}$) evaluating \textit{Perceived Usefulness}, \textit{Ease of Use}, \textit{Output Correctness}, and \textit{Reliance} on a 7-point scale through 11 questions. Participants tested the online prototype on a manuscript and then reported their experience.
Ten participants reviewed a manuscript inside their own field and one outside it. Six participants inspect cited sources often during reviews, three sometimes, and two rarely.

Responses were overall positive. 
Participants rated \toolname as highly useful (median  \surveyPuMedian{}) and easy to learn (median \surveyCOneMed{}). 
Perceived output correctness was also high. Claim identification received a median of \surveyEOneMed{} (one disagree rating). Finding the supporting evidence and verdict generation received both median ratings of \surveyETwoMed{}, with no strongly agree ratings for either.
At the same time, respondents remained cautious about relying on automated judgments. When asked whether they would accept a \textit{Support} verdict without review, median agreement was relatively low (\surveyEFourMed{}, corresponding to 3, “Disagree”). Conversely, when asked whether they would re-check a \textit{Not Supported} verdict, median agreement was high (\surveyEFiveMed{}).

A free-form question asked for improvement suggestions. In the following we provide a summary of the main improvement suggestions:
\begin{itemize}
    \item Claim identification. Extracted claims sometimes include unrelated preceding sentences. Respondents also noted missed implicit citances and lists interpreted item by item rather than collectively.
    \item Retrieval. In some paywalled cases, only the publicly visible first pages were retrieved, leading to \textit{Not Supported} rather than an inaccessible-source warning.
    \item Verdict. Respondents frequently reported \textit{Partially Supported} verdicts triggered by minor terminology differences. Some explanations also relied on manuscript context outside the checked claim. One respondent would have marked about one third of the partial-support cases as Supported.
\end{itemize}

\section{Discussion}
Due to space restrictions, we only discuss the evaluation results and the barriers to real-world adoption.
\paragraph{Evaluation Results}
No available dataset evaluates the complete workflow targeted by \toolname, which motivated our stage-wise evaluation across complementary datasets. Consequently, the reported benchmark scores should be read as stage-specific indicators. These datasets are repurposed to evaluate specific components of our pipeline rather than reproduced under their original benchmark settings. Consequently, the reported stage scores should be interpreted as indicators of performance within our evaluation setup and are not directly comparable with published results on the respective datasets. Our preliminary eight-manuscript study complements these stage-wise evaluations with an end-to-end perspective, although a broader, jointly annotated benchmark is required for stronger generalization.

\paragraph{Barriers to Real-World Adoption}
The widespread adoption and usage in real review processes is currently challenging, if not impossible, mostly due to socio-economic and not technical reasons.

First, access to full-text manuscripts is, among others, shaped by publication date and the research domain and related sharing practices. 
While open-access mandates and preprint culture are gradually reshaping how research is disseminated, this shift is a slow, long-term process whose pace varies substantially across disciplines (e.g., openness is far more established in computer science than in medicine or the humanities), and it does little to improve access to the vast body of literature published before such practices took hold.

Second, copyright constrains reuse. Some full texts sit in open-access databases (e.g., ACM's Digital Library), but these remain a minority of academic publishing. Preprint servers often make paywalled work accessible, yet preprints may only be used cautiously, as they are not the final version, and copyright agreements often restrict their usage. Per-publisher agreements are legally cleanest but do not scale, especially without institutional backing.

Third, institutional structures are required to operationalize an approach like ours at scale. To handle the financing of the LLM inference costs at scale. To help navigate the copyright issues discussed above, as publishers or consortia are far better positioned than individual researchers to negotiate processing rights. And to support data-privacy handling regarding unpublished intellectual property, ensuring manuscripts are not exposed to unwarranted retention or misuse by third-party LLM providers.

Fourth, the peer-review community needs a cultural willingness to integrate and trust AI-assisted verification tools without over-relying on their automated judgments. This requires established guidance for how such tools should complement, rather than replace, human reviewers' expertise.

\section{Threats to Validity}
\label{sec:threats}

Our evaluation and prototype design are subject to several limitations. 
First, regarding \textbf{construct validity}, the public datasets used for evaluation (e.g., Citation-Integrity, SciFact) have label definitions that do not map perfectly to our four-class verdict. For example, Citation-Integrity's \emph{Refuted} class groups direct contradiction alongside softer issues such as oversimplification and weak substantiation. Consequently, our system often predicts a cautious \emph{Partially Supported} for these cases, which penalizes the exact match accuracy despite being practically useful for reviewers.

Second, regarding \textbf{external validity}, our end-to-end evaluation on manuscripts is limited to a sample of \annotReports{} papers assessed by a single annotator. While this confirms the system's viability, the lack of inter-annotator reliability and the modest sample size limit our ability to generalize the \annotStageFourFourAcc{} accuracy across multiple research domains. Furthermore, the dataset predominantly features well-supported claims, meaning our evaluation on genuine \emph{Not Supported} citations relies heavily on the curated Citation-Integrity benchmark. Finally, the open-access resolution chain currently fails to retrieve roughly \fpeval{round((1 - \drFiveResolution{}) * 100, 0)}\% of citations due to paywalls or lack of identifiers, bounding  \toolname's effectiveness by the open-access landscape of the respective field.

\section{Conclusion}
\label{sec:conclusion}

We presented \textbf{\toolname}, a semi-automated, citation-bounded reference claim verification tool for scientific manuscripts. \toolname automates citation extraction, multi-source open-access reference retrieval, full-text evidence localization, and verdict prediction. Quantitative evaluation demonstrates strong claim detection, effective reference retrieval open-access resolution), full-paper evidence localization on whole cited papers, and \toPercent{\annotStageFourFourAcc{}} verdict agreement on real-world manuscripts within a median runtime of \drOneRuntimeMedianMin{} minutes, of which claim extraction and verdict generation take at most \drOneExtractVerdictMaxMin{} minutes.

\textbf{Ongoing and Future Work}: Ongoing work focuses on improving citation extraction across superscript and author-year formats and refining verdict rubrics for subtle overstatements. 
Future deployment depends less on further prototype functionality than on institutional partnerships. Rights-cleared access to full texts, privacy-preserving inference arrangements, sustainable cost allocation, and review-platform integration that keeps human reviewers in control. We therefore plan a live case study with an academic venue or institutional partner to evaluate usability, reviewer trust, and the practical impact of \toolname{} under these conditions.

\section*{Data Availability}
Code and data for the evaluation are available in the online material~\cite{online_material}.

\section*{GenAI Usage}
During the preparation of this work, the authors used the models Opus 4.8 and Opus 5, Gemini 3.1 Pro and Gemini 3.6 Flash for language editing and code generation, taking full responsibility for the final content.

\section*{Acknowledgements}
This work was funded by the KKS foundation through the SERT Research Profile project (research profile grant 2018/010) at Blekinge Institute of Technology.

\balance
\bibliographystyle{ACM-Reference-Format}
\bibliography{references}

\end{document}